# VISUALIZING THE INVISIBLE USING POLARISATION OBSERVATIONS

BY JO-ANNE C. BROWN, JEROEN M. STIL, AND TOM L. LANDECKER

It has been known for centuries that the Earth has a magnetic field. The idea that the Galaxy has a magnetic field and that that field might play an important role in the physics of the Galaxy is more recent, dating back almost 60 years. Fermi [1] proposed that cosmic rays may be generated outside the solar system as a result of sufficiently energetic particles colliding with moving irregularities in an interstellar magnetic field. In his view, this magnetic field would not only be a generator of cosmic rays, but also a containment factor to prevent the rays from escaping the Galaxy. Indeed, it is now believed that magnetic fields and cosmic rays contribute to the vertical support of the gas in the Galaxy [2].

In addition to playing a significant role in pressure balance, magnetic fields play an essential role in star formation, by inhibiting gravitational collapse of interstellar clouds – primary star formation regions – and by removing prestellar angular momentum [3]. Consequently, magnetic fields directly affect the distribution of stars. It is also believed that magnetic fields influence galaxy formation and evolution by causing large density fluctuations which result in structures within a galaxy [4].

*In situ* measurements of interstellar magnetic fields are not yet possible. Even the NASA Voyager Space probes, launched in 1977, have only just reached the Heliosheath [5], and are not expected to reach the interstellar medium (ISM) until 2013 at earliest. Furthermore, signatures of interstellar fields are only indirectly observable through *polarisation* observations. The majority of astronomical observations are done in total intensity, much like that of a standard photograph, thus rendering magnetic fields 'invisible' in most astronomical data. Consequently, magnetic fields have been either largely ignored in astronomy or have been used as a scape-goat for otherwise unexplained phenomena.

It has only been in the last 30 years that magnetic field observations outside our local arm have made advances, with some of the most significant steps being led by Canadians and Canadian instrumentation. For example, in almost stereotypical form, a little Canadian facility, built and run on a shoe-string budget (by any 'large-scale' facility standard) revolutionized polarisation observations with a method that would be emulated worldwide. That little facility is the Synthesis Telescope of the Dominion Radio Astrophysical Observatory (DRAO), operated by the Herzberg Institute for Astrophysics of the National Research Council. The array, built amid the mountains of the Okanagan Valley, in south-central British Columbia, consists of seven antennas in a linear east-west configuration. Three of the antennas rest on railway tracks, allowing for variable baselines (ie. pairs of correlating antennas), while the remaining four are fixed. The antennas were salvaged from various locations: two were surplus moon-radar antennas, two were surplus troposcatter antennas, two came from Five Colleges Radio Astronomy Observatory in Massachusetts, and one came from Texas where it had been used for solar radio astronomy. In order to meet weight-limit requirements at the focus, the feed horns were fashioned at a lampshade factory and the waveguides were constructed of irrigation piping. To pressurise the feed horns with dry air, aquarium pumps and canning jars are used. Yet, with clever engineering, the telescope is able to obtain high sensitivities and resolution, producing some of the best radio images in the world [6].

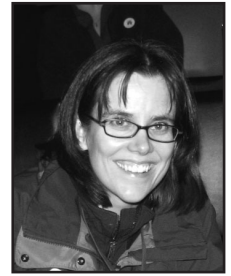

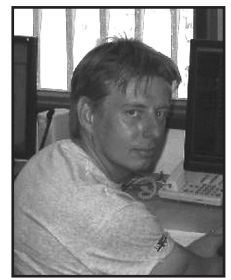

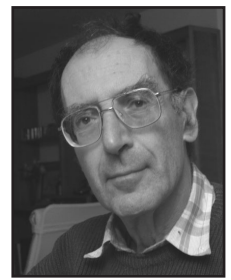

**SUMMARY**

**An electromagnetic wave can be uniquely characterized by the four Stokes parameters: *I, Q, U,* and *V*. Typical observations in astronomy rely solely on total intensity measurements of the incoming radiation (Stokes I). However, a significant amount of information both about the emitting region and the propagation path is carried in the remaining Stokes parameters. These data provide a means to observe parts of the interstellar medium which remain invisible in Stokes *I*, at any wavelength. For example, when an electromagnetic wave propagates through a region containing free electrons and a magnetic field, the plane of polarisation of the wave will rotate - an effect recorded only in Stokes *Q* and *U*. The interstellar medium of the Galaxy is such a region, containing free electrons (observed as H II) and a magnetic field of a few microgauss. By imaging in Stokes *Q* and *U* we are able to observe signatures of magnetic field perturbations from the small scale (tens of pc) to the large scale (kpc). In this paper, we review the status of Canadian polarisation studies of cosmic magnetic fields and discuss the leading role Canada is playing in polarisation studies around the world.**

J.C. Brown <jocat@ras.ucalgary.ca>, J.M. Stil, Centre for Radio Astronomy, University of Calgary, Calgary, AB T2N 1N4

and

T.L. Landecker, Dominion Radio Astrophysical Observatory, National Research Council Canada, Penticton, B.C. V2A 6J9

This observatory was the primary instrument of the Canadian Galactic Plane Survey (CGPS [7]), a multi-facility project designed to image a section of the Galactic plane at multiple wavelengths and full polarisation. After demonstrated success in both observation technique and discoveries, the CGPS was then expanded into the International Galactic Plane Survey, which included the Southern Galactic Plane Survey (SGPS [8,9]) with observations taken from the Australia Telecope Compact Array (ATCA). The survey areas for both projects are shown in Figure 1. While there have been many significant discoveries made through these surveys, the ones we will focus on in this paper relate to polarisation observations in general, and magnetic fields in particular. We will also discuss how our knowledge gained with these projects has prepared us, as Canadians, for the international stage with large upcoming projects like the Square Kilometer Array.

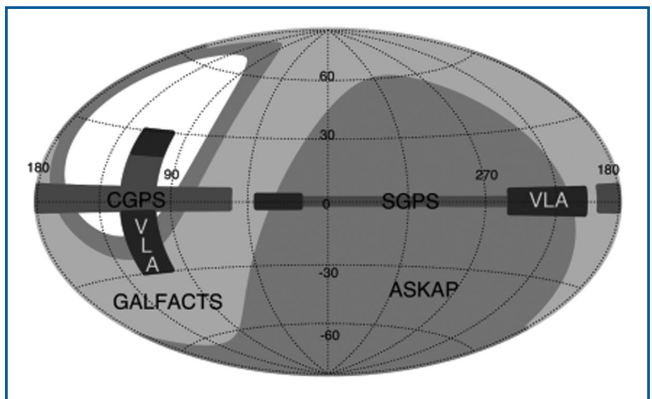

Fig. 1   All sky view (in Galactic coordinates) of past, present, and future radio observation surveys with significant Canadian involvement. CGPS: Canadian Galactic Plane Survey (PI at UofC); SGPS: Southern Galactic Plane Survey (UofC participation); VLA: Very Large Array observations (Co-PI at UofC) ; GALFACTS: Galactic Arecibo L-band Feed Array Continuum Transit Survey (PI at UofC); ASKAP: Australian Square Kilometer Array Pathfinder (Canada is a formally recognized partner). The Global Magneto-Ionic Medium Survey (GMIMS: PI at NRC) will cover the entire sky.

## TECHNIQUES FOR OBSERVING MAGNETIC FIELDS

The interstellar medium consists of several basic constituents: atomic, molecular and ionized gas, dust, cosmic rays and magnetic fields [10]. Most of the constituents have some form of observable radiation, and may be observed directly at the appropriate wavelength. Such observations are referred to as 'total intensity' or 'Stokes $I$' observations. Unlike these other constituents, magnetic fields themselves do not radiate, and consequently, cannot be observed directly. However, they can affect the sources of radiation or the radiation directly, given the right conditions. The signature of these effects shows up in the other Stokes parameters, primarily $U$ and $Q$, as we discuss below.

**Polarisation and Stokes Parameters**

The concepts of polarisation and Stokes parameters are often minimized or bypassed altogether in the undergraduate curriculum in physics, though they are of fundamental importance in electrical engineering, where radio astronomy has its roots. In fact, all man-made electromagnetic signals are polarised because antennas are made of wires which channel the electron flow, imposing a preferred direction on the emitted radiation. The only signals that can be unpolarised are natural signals.

We briefly review the relevant aspects of polarisation and the Stokes parameters in this section, and discuss how they are exploited to remotely study magnetic fields in the subsequent sections. Additional details may be found in Refs [11] and [12].

Electromagnetic waves are transverse, meaning that their oscillations are perpendicular to their direction of propagation. If the wave vector and wave electric field define a plane that does not change as the wave propagates, then the wave is *linearly polarised*, since the wave is seen to define a line when viewed along the direction of propagation.

If we consider two orthogonal, linearly polarised electromagnetic waves of the same frequency travelling in the $z$ direction, with the first polarised in the $x$ direction ($x$-$z$ plane), and the second in the $y$ direction ($y$-$z$ plane), the electric fields of the two waves may be described by the following equations:

$$E_x = E_1 \cos(kz - \omega t) \quad (1)$$

$$E_y = E_2 \cos(kz - \omega t - \delta) \quad (2)$$

where $k$ is the wavenumber, $\omega$ is the frequency, $t$ is time, and $\delta$ is the phase offset between the two waves ($\delta = \delta_x - \delta_y$). The detected wave will be the vector sum of these two individual waves such that $\mathbf{E} = E_x \hat{\mathbf{i}} + E_y \hat{\mathbf{j}}$. At $z = 0$, the components of $\mathbf{E}$ may be reduced to:

$$E_x = E_1 \cos(\omega t) \quad (3)$$

$$E_y = E_2 \cos(\omega t + \delta) \quad (4)$$

Combining equation 3 and equation 4 results in the equation of an ellipse:

$$1 = aE_x^2 - bE_xE_y + cE_y^2 \quad (5)$$

where

$$a = \frac{1}{E_1^2 \sin^2 \delta}, \quad b = \frac{2\cos\delta}{E_1 E_2 \sin^2 \delta}, \quad c = \frac{1}{E_2^2 \sin^2 \delta}. \quad (6)$$

This equation describes the locus of points traced out by the vector $\mathbf{E}$ as it propagates. The ellipse, known as the polarisation ellipse, may be characterized by two angles, $\tau$ and $\varepsilon$, as illustrated in Figure 2.

Critical to our work is the angle $\tau$, known as the polarisation angle. It gives a measure of inclination of the ellipse with respect to the $x$ axis [1] and is defined within the limits of $0° \leq \tau < 180°$, since $\tau = 180°$ is indistinguishable from $\tau = 0°$. The second angle, $\varepsilon$, is in essence a measure of the ellipticity of the wave. Its value is determined by the arc-cotangent of the ratio of the major axis (OA) to the minor axis (OB) and is defined within the limits $-45° \leq \varepsilon \leq +45°$. Negative values of $\varepsilon$ correspond to 'right-handed' (or right-elliptically polarised) waves, where **E** moves counterclockwise when viewed travelling towards the observer, while positive values of $\varepsilon$ correspond to 'left-handed' (or left-elliptically polarised) waves, where **E** moves clockwise as viewed from the same vantage point. This definition of handedness is the IEEE convention. It is the standard in radio astronomy and is consistent with the well known 'right hand rule'. [2]

---
1. In astronomy, the x axis is defined as 'sky North'.

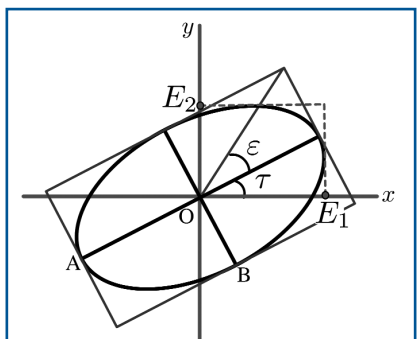

Fig. 2 The polarisation ellipse (after figure 4.5 in Ref. [11]). $E_1$ and $E_2$ are the magnitudes of two monochromatic waves of identical frequency, polarised in the $x$ and $y$ directions respectively. The vector sum of these two waves traces out an ellipse with a polarisation angle of $\tau$ (see text for more details).

Depending on the properties of $E_1$, $E_2$ and $\delta$, the polarisation ellipse will take on different forms. In general, waves with $0° < \delta < 180°$ will be left elliptically polarised whereas waves with $180° < \delta < 360°$ ($-180° < \delta < 0°$) will be right elliptically polarised. If $\delta = 0°$ or $\delta = 180°$, the wave will be linearly polarised. In cases where $E_1$, $E_2$ and $\delta$ are such that elliptical polarisation results, the wave can be thought of as having some component of linear polarisation, and some component of circular polarisation.

The state of polarisation represented by a polarisation ellipse may be described mathematically by the four Stokes parameters. Introduced in 1852, they are

$$I = (E_1^2 + E_2^2)/Z \qquad (7)$$
$$Q = (E_1^2 - E_2^2)/Z \quad = I \cos 2\varepsilon \cos 2\tau \qquad (8)$$
$$U = (2E_1 E_2 \cos \delta)/Z \quad = I \cos 2\varepsilon \sin 2\tau \qquad (9)$$
$$V = (2E_1 E_2 \sin \delta)/Z \quad = I \sin 2\varepsilon \qquad (10)$$

where $Z$ is the impedance of the medium [11]. Stokes $I$ is the total intensity of the wave, Stokes $Q$ and $U$ are measures of the linear polarisation of the wave, and Stokes $V$ is a measure of the circular polarisation of the wave.

With these two formulations of the Stokes parameters (the relationship between the two may be found in Ref. [11]), it is easy to see that the first definition allows for straight-forward measurements of the parameters by an antenna with a given impedance $Z$ designed to measure linear polarisation on two orthogonal axes. Once the Stokes parameters have been determined, the second formulation allows for the calculation of $\tau$ and $\varepsilon$. In particular, we note that

$$\tau = \frac{1}{2}\tan^{-1}\frac{U}{Q} \qquad (11)$$

The above discussion dealt with a completely polarised or monochromatic wave, where $E_1$, $E_2$ and $\delta$ are constant. In general, emissions from celestial radio sources extend over a wide range of frequencies. Within any finite range of frequencies detected by a receiver, the wave will consist of a superposition of a large number of statistically independent waves with a variety of polarisations. Therefore, $E_1$, $E_2$ and $\delta$ will be detected as having time dependence, and the Stokes parameters will use the time-averages of these values.

In the pure, monochromatic case, $I^2 = Q^2 + U^2 + V^2$. With multiple wave fronts averaged together, it is possible to have $I^2 \geq Q^2 + U^2 + V^2$. For a completely unpolarised wave, $Q = U = V = 0$. The degree or fraction of polarisation is defined as

$$d_p = \frac{\text{polarised power}}{\text{total power}} = \frac{\sqrt{Q^2 + U^2 + V^2}}{I} \qquad (12)$$

where $0 \leq d_p \leq 1$. Thus, $d_p = 1$ for a completely polarised wave, while $d_p = 0$ for a completely unpolarised wave.

In the interstellar medium, most of the polarised radiation we observe at radio wavelengths comes from synchrotron emission (see section below), and is consequently linearly polarised. Therefore, when we talk about *polarised intensity*, we are really talking about *linearly* polarised intensity, defined as

$$PI = \sqrt{U^2 + Q^2}. \qquad (13)$$

Figure 3 shows CGPS Stokes $I$, linear polarised intensity (PI) and polarisation angle ($\tau$) images of a small part of the Galactic plane. The CGPS images are unique because of their image fidelity, and the inclusion of short spacing information over a large area of the Galactic plane. The most striking aspect of Figure 3 is the wealth of structure in polarisation angle and intensity, which is not seen in Stokes $I$. In fact, it is rare to find a counterpart of polarised structure in the total intensity images. An exception is the low polarised intensity in the upper right corner, associated with depolarisation by tangled magnetic fields in a low-density halo around the bright HII (ionized hydrogen) region W4 seen in the Stokes $I$ image [13].

**Faraday Rotation**

When a linearly polarised electromagnetic wave propagates through a region of free electrons permeated by a magnetic field (e.g. a magnetized plasma such as the interstellar medium), its plane of polarisation will rotate. This phenomenon is known as Faraday rotation. Faraday rotation can be understood as a consequence of *birefringence*, where the magnetised plas-ma has two different indices of refraction corresponding to two different states of incident polarisation [16].

---

2. Under the classical physics convention, the handedness definition is reversed.

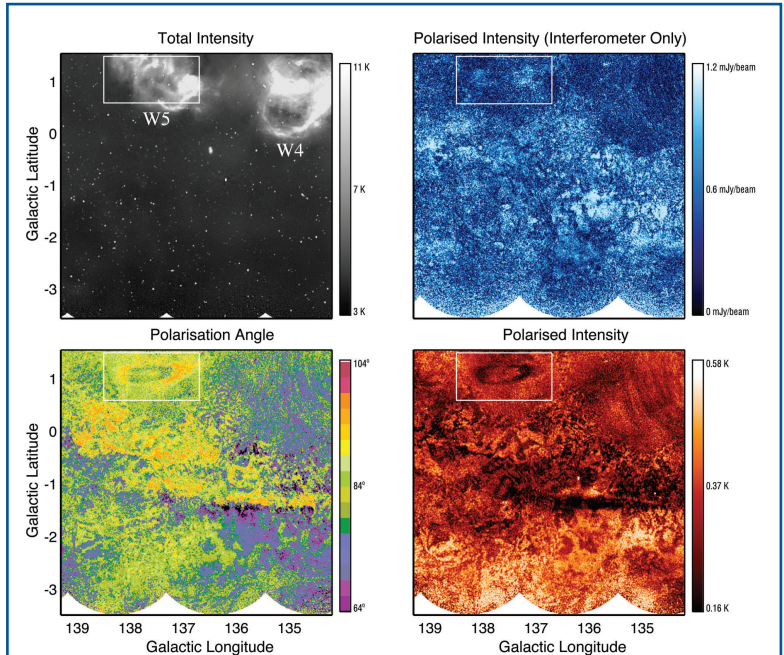

Fig. 3  Mosaic MX1 of the CGPS in total intensity (Stokes *I* ), total linear polarised intensity (with and without the zero-spacing data), and polarisation angle. "Zero spacings" or single-dish data have been added to the interferometric data for all images except the 'interferometer only' polarised intensity. For the purpose of this illustration, the polarisation angle has been shifted by 90º. As shown, the polarisation angles have a relatively narrow distribution, which is a consequence of the local large-scale magnetic field. This plot demonstrates how different a region can look in polarisation compared to Stokes *I*, as well how the addition of zero-spacing data can affect the polarisation images themselves. W4 and W5 are ionized hydrogen (HII) regions first identified by Westerhout [14]. The box highlights the Faraday rotation feature first identified by Gray *et al.* [15].

For a linearly polarised wave, the birefringence is with respect to right and left circularly polarised waves (an alternative to the linearly polarised basis set used above). The birefringence will slow one of the circularly polarised waves with respect to the other, resulting in a rotation of their sum, the linearly polarised wave (e.g. Ref. [17]).

To calculate the indices of refraction for a cosmic magnetised plasma, the *quasi-longitudinal* (QL) approximation is invoked [18]. The validity of the QL approximation depends on how closely the direction of propagation of a wave ($\hat{\mathbf{k}}$) coincides with the field direction ($\hat{\mathbf{b}}$), defined as $\theta = \cos^{-1}(\hat{\mathbf{b}} \cdot \hat{\mathbf{k}})$, and on the electron density and the collision frequency. As $\theta$ approaches 90º a linearly polarised wave will aquire some ellipticity (known as the Cotton-Mouton effect), instead of simply rotating as it would in the QL regime.

The QL approximation is valid if [19]

$$\frac{\Omega}{\omega} \sec \theta \ll 1 \quad (14)$$

and if

$$\frac{\omega_p^2}{\omega^2} \ll 1, \quad (15)$$

where $\Omega$ is the cyclotron frequency, and $\omega_p$ is the plasma frequency. If the total strength of the Galactic magnetic field is, on average, roughly $B_o = 10\ \mu G$ ($10^{-9}$ T), the electron cyclotron frequency is

$$\Omega = \frac{eB_o}{m_e} \quad (16)$$
$$= 175.63\ \text{rad/s}$$

where $e$ and $m_e$ are the electron charge and mass, respectively. With $(\Omega/\omega) \sec \theta = \frac{1}{1000}$, and $\omega = 2\pi \times$ 1420 MHz, the QL approximation holds for $\theta <$ 89.99887º. Similarly, for equation 15, using $n_e =$ 1 cm$^{-3}$ ($10^6$ m$^{-3}$), the plasma frequency is

$$\omega_p = \left( \frac{n_e e^2}{\varepsilon_o m_e} \right)^{\frac{1}{2}} \quad (17)$$
$$= 56 \times 10^3\ \text{rad/s}$$

As a result, $\omega_p^2/\omega^2 = 4 \times 10^{-11} \ll 1$. The fact that the QL approximation holds for such a large range of angles at radio wavelengths often leads one to forget that it *is* an approximation that must be verified depending on the application and region of the ISM being explored.

Invoking the QL approximation and using the resultant indices of refraction for circularly polarised waves in the ISM plasma (see Ref. [10]), the amount of rotation a radio wave will acquire, $\Psi$, is given by

$$\Psi = \lambda^2 (0.812 \int n_e \mathbf{B} \cdot d\mathbf{l})\ [\text{rad}] \quad (18)$$
$$= \lambda^2\ \text{RM}$$

where $\lambda$ is the wavelength in units of m, $n_e$ is the electron density in units of cm$^{-3}$, **B** is the magnetic field in units of $\mu$G, $d\mathbf{l}$ is the incremental pathlength in units of pc, and RM is the **rotation measure**:

$$\text{RM} = 0.812 \int n_e \mathbf{B} \cdot d\mathbf{l}\ [\text{rad m}^{-2}] \quad (19)$$

It is important to recognize the significance of three key elements of equation 18. First, it is wavelength dependent. As a result, waves of different frequency will experience different amounts of rotation through the same plasma. Second, the effect of Faraday rotation is weighted by the electron density; higher electron densities will result in greater rotation. Finally, it is the *direction* of the line-of-sight component of the magnetic field ($B_\parallel$) that determines the *sign* of the rotation measure. Since the path length is defined to be *from* the source *to* the receiver, (ie. the telescope on Earth), a magnetic field with $B_\parallel$ directed towards us results in a positive RM, while a magnetic field with $B_\parallel$ directed away from us results in a negative RM.

With this in mind, if we assume that at all wavelengths, the polarised emission from a given source is emitted at the same polarisation angle, $\tau_o$, and that the radiation is only affected by Faraday rotation, then the detected polarisation angle, $\tau$ at a given wavelength $\lambda$, will be given by

$$\tau = \tau_o + \lambda^2 \, \text{RM}. \tag{20}$$

Since this relationship is linear, measurements of $\tau$ at multiple wavelengths can determine the RM for a given source as the slope of the graph of $\tau$ versus $\lambda^2$.

The ease with which RMs can be determined, coupled with the significance of the sign of the RM, makes RM measurement a powerful tool for probing the ISM magnetic field. Pulsars and extragalactic sources (EGS) are sources of linearly polarised radiation and are often used as compact (or point source) probes of the Galactic magnetic field. Since it is possible to estimate the distance to these sources, and if we know something of the electron density along their lines-of-sight, then we can work backwards to estimate what the magnetic field must look like along their particular plumb-lines. Subsequently, the goal for observations is to measure RMs for the highest density of sources possible, allowing for the most accurate reconstruction of the intervening field.

Prior to the CGPS, observations of multiple polarisation angles for EGS were done at widely separated wavelengths, often at different times, and sometimes even at different facilities. Consequently, there was uncertainty as how to 'unwrap' the polarisation angles in order to determine the correct RM (e.g. Refs. [20,21]). DRAO was the first facility to do polarisation measurements at 4 wavelengths sufficiently close together so that the ambiguity in RM calculations was removed [22]. This technique was emulated at the ATCA for the SGPS. Instead of 4 bands, the SGPS had 12 bands, improving on the technique initiated by the CGPS.

Not only did the CGPS set the standard for observation techniques, it also set the standard for EGS RM source density. Prior to the CGPS, Broten et al. [23] had compiled a catalog of high quality EGS RM measurements. With 674 sources in the catalog, the majority of which were out of the Galactic plane, the RM density was roughly 1 source per 60 square degrees. The CGPS produced (and continues to produce) RMs at a density of 1 source per square degree, resulting in significantly more reliable conclusions about the magnetic field than previously possible. The SGPS source density is slightly lower than the CGPS, at 1 source per 2 square degrees, as a result of depolarisation through the inner Galaxy. However, it must be noted that prior to the survey, there was only 1 published EGS RM in the entire SGPS region.

**Synchrotron emission**

The classic source of radiation is accelerating charges. Therefore, charged particles moving in the presence of a magnetic field will undergo acceleration through the Lorentz force, and subsequently radiate. If the particles are moving at relativistic speeds, this radiation is called synchrotron radiation.

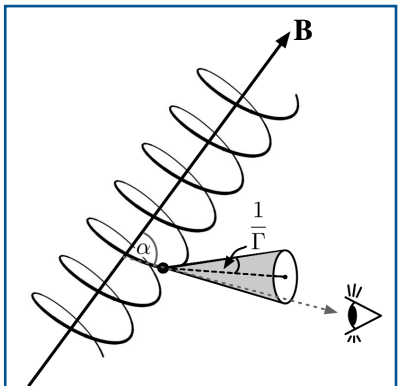

Fig. 4  Diagram of a relativistic electron with Lorentz factor $\Gamma$ spiraling in a uniform magnetic field. The electron emits synchrotron radiation with a high degree of linear polarisation in a narrow cone directed along the instantaneous velocity of the electron.

Radio emission at decimeter wavelengths from our Galaxy is dominated by synchrotron emission from relativistic electrons with Lorentz factor $\Gamma \geq 10^5$ in a magnetic field. Relativistic particles from interstellar space, named cosmic rays for historic reasons, were first observed on Earth in balloon experiments by Hess in 1912.

Figure 4 shows a relativistic electron on a helical path in a uniform magnetic field due to the Lorentz force. In the observer's rest frame, the radiation emitted by the accelerated electron is emitted in a narrow cone along the electron's velocity vector as a result of the relativistic beaming effect (e.g. Ref. [24]). The opening angle of the emission cone is $1/\Gamma$, much narrower than shown in Figure 4. Consequently, the observer sees only emission from those electrons that have an instantaneous velocity directed towards the observer (pitch angle $\alpha$). Furthermore, electrons traveling parallel to the field will not be accelerated, and will therefore not radiate. Thus, the amount of synchrotron emission observed depends on the presence of a magnetic field component perpendicular to the line of sight, $B_\perp$. Optically thin synchrotron emission $\epsilon_s$ at frequency $\nu$ of an ensemble of relativistic electrons with a power law energy spectrum $N(E) \sim E^\gamma$ in a uniform magnetic field depends on the magnetic field component in the plane of the sky $B_\perp$ according to

$$\epsilon_s \sim B_\perp^{(\gamma+1)/2} \nu^{-(\gamma-1)/2} \tag{21}$$

Synchrotron emission provides information about the slope of the energy spectrum of the electrons, and an estimate of the strength of the magnetic field if assumptions are made about the volume of the source and the energy density of cosmic rays. The minimum combined energy density of cosmic ray particles and magnetic field for an observed source is similar to the equipartition energy density of the magnetic field. Magnetic field estimates from the brightness of synchrotron emission assume this minimum energy condition, without clear justification that the minimum energy density condition or equipartition apply.

Synchrotron emission from a region with a uniform magnetic field has a theoretical limit of $\sim 70\%$ [25], with the plane of polarisation perpendicular to the direction of $B_\perp$ in the plane of the sky. Polarisation of synchrotron emission therefore gives information on the magnetic field component perpendicular to the line of sight, while Faraday rotation gives information on the magnetic field component along the line of sight.

In practice, the observed emission is integrated over large regions in space with a complicated magnetic field structure and an ionized plasma present in addition to the relativistic electrons. At decimetre wavelengths the integrated emission is usually much less polarised than the theoretical 70%. In this wavelength range thermal emission is mostly fainter than synchrotron emission unless the line of sight crosses a denser HII region, ionized by massive stars. Nevertheless, it is very common for a plasma structure to cause significant Faraday rotation, yet to be undetectable by its thermal emission. Faraday rotation alone cannot alter the amplitude of the polarised signal, but a number of physical and instrumental effects often reduce the fractional polarisation of the recorded signal [26,27]. Differential Faraday rotation (or depth depolarisation) occurs when synchrotron emission generated at different depths along the line of sight suffers different rotation, and vector averaging reduces the observable polarised intensity. Beam depolarisation occurs when many turbulent cells and/or large RM gradients exist within the beam of the telescope, again leading to vector averaging. On most angular scales these effects create structures in polarised intensity, and particularly in polarsation angle, that have no counterpart in total intensity. Since these effects are most pronounced at decimetre wavelengths, that wavelength regime provides the best data for studying the magnetic field configuration within the interstellar medium.

The detection of linear polarisation in the extended Galactic radio emission [28,29] provided crucial evidence in establishing the synchrotron mechanism that makes the Milky Way a strong radio source. The apparent potential of polarisation observations to reveal the Galactic magnetic field led to efforts to map polarised emission over wide areas of the sky. The best of these datasets is that of Brouw and Spoelstra [30] who presented data from the Dwingeloo 25-m Telescope for much of the Northern sky at four frequencies between 408 and 1411 MHz. Angular resolution ranged from 2º to 36′, but the sampling was far from complete. A major Canadian contribution to this field is the 1.4 GHz polarisation survey of Wolleben et al. [31] made with the DRAO 26-m Telescope. The northern sky was mapped down to declination -30º with 200 times more data points than the Dwingeloo data and five times better sensitivity, but based on the absolute calibration of Brouw and Spoelstra. These new data have played an important role in the cosmology industry, by providing a significant counterpoint to the Wilkinson Microwave Anisotropy Probe (WMAP) 23 GHz data [32] as well as providing a clearer understanding of the polarised features observed (e.g. Ref. [33]).

**MAGNETIC STRUCTURE IN THE LOCAL ISM**

Cosmic ray electrons in the Galaxy emit synchrotron emission that is observed as a featureless glow across the sky with a tendency to be brighter near the Galactic plane. In addition to this smooth synchrotron background we see synchrotron emission from specific objects, mainly supernova remnants. The diffuse synchrotron emission originates from a large volume in space. It takes a substantial line of sight distance to build up detectable synchrotron emission, because the highly relativistic electrons that emit this radiation are so rare. The same volume of space is littered with plasma structures that give rise to Faraday rotation and depolarisation effects described above. It comes as no surprise that the interpretation of polarisation of diffuse Galactic emission is very difficult. However, it is also the only way we can observe most of the magneto-ionized interstellar medium.

Wieringa et al. [34] reported structures in polarised radio emission at 327 MHz that had no counterpart in total intensity. It was soon realized that these structures were small-scale modulations in the polarisation angle of Galactic synchrotron emission that were detectable by the radio interferometer. The structure in polarisation angle gave rise to structure in the Stokes $Q$ and $U$ images, even though the interferometer did not detect the smooth emission in total intensity because of the so-called missing short spacings; Faraday rotation effects tend to break large structures into structure on smaller angular scales. Nevertheless, inclusion of data from a large single-dish radio telescope provides information on the largest structures that an interferometer cannot detect and has a dramatic effect on the polarisation images. Some polarised features are observed to change significantly with the inclusion of the large-scale polarised emission not seen by the interferometer (see Figure 3).

The study of structure in diffuse polarised emission really took off with the CGPS. The polarisation survey of the CGPS at 1.4 GHz [35] marks a major advance in polarization observations. Data from the DRAO Synthesis Telescope, the Effelsberg 100-m Telscope, and the DRAO 26-m Telescope have been combined to give accurate representation of all structures down to the resolution limit of ~1′. With $1.7 \times 10^7$ independent data points, this is the largest polarisation survey made to date, and the most extensive dataset to combine single-antenna and aperture-synthesis data. For this survey, new techniques were developed to correct for instrumental polarisation across the field of view [36,37] and independent calibrations of the three datasets were carefully compared.

An origin in Faraday rotation means automatically that the polarised "objects" will usually not look like things seen in other wavelengths. Since the polarised sky is almost entirely new, the first task is to classify the various polarisation features, though it is often difficult to draw a boundary around such objects. Early results from the survey have identified two enigmatic Faraday rotation features [15,38], with a similar object reported by Haverkom et al. [39]. Polarisation features associated with known objects have also been identified, including that associated with a planetary nebula [40] and a stellar wind bubble [41].

Figure 3 shows the lenticular feature of Gray et al. [15] which resides in the direction of the HII region W5. The longest size of this feature is approximately two times the diameter of the full moon. While visible in polarisation angle, it was not visible in polsarised intensity prior to the addition of the zero-spacings, nor is there any obvious counterpart in Stokes $I$. While the origin of such "polarisation lenses" is still not clear, it is likely

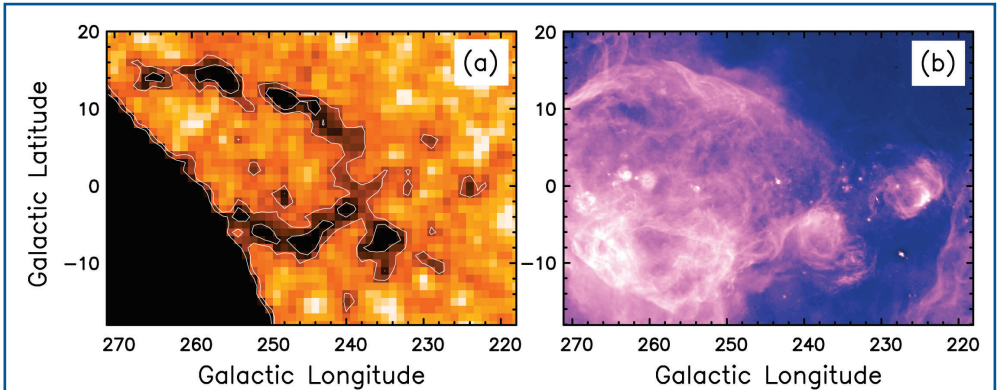

Fig. 5 (a) Number density of polarised sources in the NVSS from Stil and Taylor [42]. Contours are drawn at 2.7 and 4.0 sources per square degree. The black area in the lower left corner is below the southern declination limit of the NVSS. (b) Hα intensity in the same region from the SHASSA survey [43].

either a concentration of electron density or a magnetic field structure.

Looking at the polarisation images of the CGPS in general, it is almost impossible to find the Galactic plane: the polarised signals seem to continue undimmed to the southern and northern limits of the data. This is in sharp distinction to the total-intensity distribution and the distributions of other ISM tracers such as dust. The simplest interpretation is that the polarised emission that we are seeing at 1.4 GHz is generated in nearby volumes of the ISM. This is consistent with the concept of the *polarisation horizon* introduced by Uyaniker et al. [44]. The combined effects of depth depolarisation and beam depolarisation do not allow us to detect polarised emission beyond a certain distance. The distance to the polarisation horizon depends on frequency, beamwidth, and direction.

Some local structures escape detection in observations of diffuse polarised emission. Figure 5 shows a large magnetized shell in the Gum nebula, that was revealed because it depolarises extragalactic radio sources [42]. The shell is so large on the sky that the Big Dipper would fit inside it. Figure 5(a) shows the sky density of polarised extragalactic sources found in the NRAO VLA Sky Survey (NVSS; Ref. [45]). The dark ring represents a lack of polarised sources, depolarised because of strong Faraday rotation in a shell with a compressed magnetic field. The upper part of this shell is visible in the Hα emission in Figure 5(b). The lower part of the shell is clearly defined by the lack of polarised sources in Figure 5(a), but bright Hα emission that does not depolarise background sources confuses the image in Figure 5(b). Vallée and Bignell [46] first suggested the presence of a magnetised shell because of anomalously high rotation measures for some extragalactic sources.

## THE GLOBAL MAGNETIC FIELD IN OUR GALAXY

Despite its recognized importance, very little is truly known about the Galactic magnetic field. What is its source? How is it maintained? We cannot hope to answer these questions until we understand what it really looks like: where it is (and isn't!) located, what its direction and magnitude are, and how it is correlated (or uncorrelated) with the medium in which it appears to be embedded.

At one time, it was believed that the Galactic magnetic field was primordial in origin, meaning it was present as a weak 'protofield' at the time the Galaxy was formed, and it subsequently evolved and amplified as the protogalaxy contracted and rotated. The primary objection to the primordial theory is based on the time scales required to generate the observed fields existing in galaxies [47].

However, a primordial field may have served as the seed field for a Galactic dynamo [48]. By definition, a dynamo converts the energy of motion of a conductor into the energy of an electric current and a magnetic field [49]. In the Galaxy, the conducting fluid requirement for a dynamo is satisfied by the ionized interstellar gas. The differential rotation of the Galaxy could produce appropriate fluid motions that would amplify the seed field. Dynamo theory is currently favored among magnetic field theorists as it appears to be quite robust and seems to be able to provide a universal explanation of the varied field configurations observed [50].

This is where much of the observational work is focussed - identifying the topology of the field to provide adequate constraints for modeling in order to determine the most likely mode(s) of the dynamo(s) acting in the Galaxy. While some features of the Galactic magnetic field are universally accepted as facts, others remain highly contentious. We discuss both the accepted and contentious features of the Galactic magnetic field in the following sections.

### Accepted Observational Constraints

The Galactic magnetic field is usually considered to be composed of two distinct components: a smooth or uniform component, $\mathbf{B}_u$, with scale sizes on the order of a few kiloparsecs (kpc), and a turbulent or random component, $\mathbf{B}_r$, with scale sizes on the order of tens of parsecs (pc; Ref. [51]). The uniform component is observed to be concentrated in the disk [48,52], with a dominant azimuthal component, some radial component (indicating a spiral field), and a weak vertical or $z$ component. Conversely, $\mathbf{B}_r$ is believed to be isotropically distributed as integrated along the line-of-sight [53], though there is some evidence to suggest that $\mathbf{B}_r$ is correlated with $\mathbf{B}_u$ [54], and that the scale-sizes of $\mathbf{B}_r$ are significantly different between and within the spiral arms [55,56].

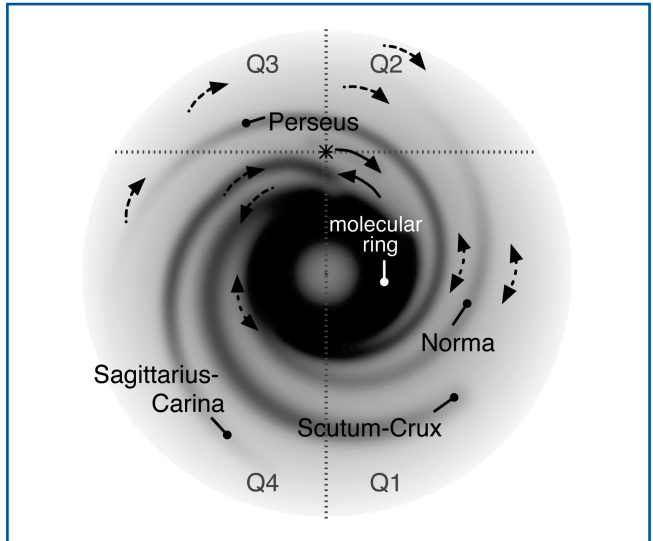

Fig. 6   The directions of the large-scale Galactic magnetic field as viewed from the North Galactic Pole. The grey scale is the CL02 model of the electron density [57]. Q1-4 indicate the four Galactic quadrants, and the asterisk indicates the location of the Sun. Solid arrows indicate universally accepted field directions; single-ended dashed arrows indicate field directions as supported by Canadian work, though not necessarily universally accepted; double-ended arrows indicate regions remaining highly debated with no recent Canadian input.

As shown in Figure 6, the field within the local arm is observed to be directed clockwise, as viewed from the North Galactic pole, with a strength of roughly 6 μG [53,58] [3]. In the first quadrant (Q1) of the Sagittarius-Carina arm, the magnetic field is unquestionably observed to be directed counter-clockwise [59,60], indicating a region of magnetic shear between the local and Sagittarius-Carina arms. Such a region is what we call a magnetic field reversal. The number and location of such magnetic field reversals are arguably the most significant factors in differentiating between likely dynamo modes.

**Controversial Observational Constraints**

The paucity of available data makes identifying magnetic field reversals exceedingly difficult. Consequently, different interpretations of similar data do occur.

Based primarily on RMs of EGS, neither Simard-Normandin and Kronberg [52] nor Vallée [61] could find any evidence for reversals beyond the solar circle. However, using a limited number of pulsar RMs along with the EGS RMs available at the time, Rand and Kulkarni [62] and Clegg et al. [63] suggested the presence of a field reversal associated with the Perseus arm, and Han et al. [64] concluded that there may be an additional reversal beyond the Perseus arm. With the significant increase in EGS RM source density of the CGPS, coupled with newer pulsar RMs, Brown et al. [22] demonstrated that a reversal is not necessary to explain the observations, thus coming full circle to the initial conclusion of no reversals beyond the solar circle.

In the inner Galaxy, studies using both pulsars and EGS RMs have produced evidence suggesting the field reverses back to a clockwise direction at R ~ 5.5 kpc for the Scutum-Crux arm [21], and perhaps switches again at R ~ 3 kpc for the Norma arm [64]. The apparent pattern of the field reversing with every arm is supported by the recent work of Weisberg et al. [65], who studied pulsar RMs primarily in Q1. Han et al. [66] has even suggested the field reverses at every arm-interam interface. Conversely, using the new SGPS EGS RMs along with the pulsar RMs, Brown et al. [67] could only find strong evidence for one reversal, and weaker evidence for a second inside the solar circle. Furthermore, the strong first reversal is seen to occur between the Sagittarius-Carina arm and Scutum-Crux arm in quadrant 4 (Q4), instead of between the local arm and Sagittarius-Carina arm as observed in Q1. This suggests the field has much less inclination (ie. it is more azimuthal) than the optical spiral arms. This is in agreement with the interpretation of the pulsar RM data made by Vallée [68] who envisaged a ring model with a reversal that passes through the Sagittarius-Carina arm around Galactic longitude = 0 (ie. the Sun - Galactic centre line).

Clearly more data are required to differentiate between these differing opinions. To that end, we recently acquired time on the VLA to fill in the gaps between the CGPS and the SGPS, as shown in Figure 1. We hope these data will be sufficient to determine the field structure in the inner Galaxy. Otherwise, we will have to wait for data from the upcoming projects described in the section on Current and Future Projects with Significant Canadian Content.

**THE MAGNETIC FIELD IN EXTERNAL SPIRAL GALAXIES**

While living *inside* of a galaxy provides unique opportunities to study galactic dynamics up close, it carries with it the inherent problem of the 'forest-for-the-trees' effect. Therefore, it is extremely beneficial if we balance observations within our own Galaxy with those of external spiral galaxies.

Observing magnetic fields in external galaxies has the advantage that it is easier to see structure from the outside, and magnetic fields can be studied in a variety of galaxies with different properties. However, different techniques must be employed than for observations of the Galactic magnetic field. Even for galaxies as close as a few Mpc, present-day telescopes cannot detect a sufficient number of background EGS to do statistical RM studies as is done in our Galaxy. To date, only two external galaxies, M31 [69] and the Large Magellanic Cloud [70], have been probed with RMs of background sources. Both galaxies show a large-scale, regular magnetic field, but no field reversals were detected.

Most observations of magnetic fields in external galaxies are limited to polarisation of synchrotron emission with a resolution up to ~10″, corresponding with 240 pc at a distance of

---

3. For perspective, the strength of the Earth's magnetic field is 0.6 G at the poles.

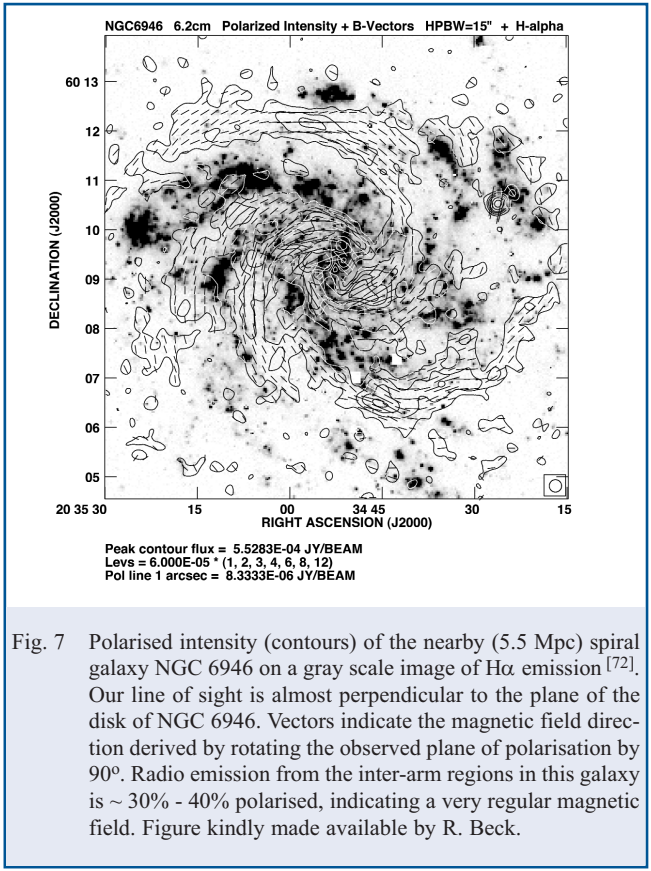

Fig. 7  Polarised intensity (contours) of the nearby (5.5 Mpc) spiral galaxy NGC 6946 on a gray scale image of Hα emission [72]. Our line of sight is almost perpendicular to the plane of the disk of NGC 6946. Vectors indicate the magnetic field direction derived by rotating the observed plane of polarisation by 90°. Radio emission from the inter-arm regions in this galaxy is ~ 30% - 40% polarised, indicating a very regular magnetic field. Figure kindly made available by R. Beck.

5 Mpc. All but the largest polarised structures identified in diffuse Galactic emission (section on Magnetic Structure in the Local ISM) would be unresolved in radio images of nearby spiral galaxies. As the resolution is also similar to the outer scale of energy injection into the interstellar medium by stars, most structures in the interstellar medium only contribute to an unresolved stochastic component of the rotation measure. This structure within the beam leads to depolarisation of the emission. Remarkably, radio emission of some galaxies is locally 30 - 40% polarised at a wavelength of 6 cm [71,72].

Images of polarised emission of spiral galaxies reveal the magnetic field on a galactic scale, projected on the plane of the sky. As in our own galaxy, the regular magnetic field is predominantly in the azimuthal direction in the plane of the disk. The direction of the regular magnetic field shows a spiral pattern similar to the optical spiral arms, but the most regular fields are found away from the spiral arms in inter-arm regions. Figure 7 shows polarised emission of the galaxy NGC 6946 [72] in relation to the spiral arms as traced by the Hα emission of massive star formation regions. The total magnetic field in the (optical) spiral arms is actually a factor ~2 *stronger* than in the highly polarised magnetic arms, but the field in the spiral arms is more tangled on scales smaller than the resolution of the image, resulting in a low degree of polarisation.

Large-scale departures from a symmetric, azimuthally oriented magnetic fields are found in some galaxies. Barred galaxies can have a magnetic field oriented along the bar. Some galaxies with an intense star burst and an associated outflow of gas perpendicular to the disk display large-scale magnetic fields perpendicular to the disk, and far into the halo, e.g. NGC 4569 [73]. Gravitational interaction with another galaxy or ram pressure interaction with an intracluster medium can deform the disk of a galaxy and its associated magnetic field [74].

Observational evidence for the evolution of magnetic fields in galaxies with cosmic time has been elusive because radio observations of spiral galaxies at cosmologically interesting distances are not possible with current instruments. Indirect evidence for substantial magnetic fields in normal galaxies comes from the association of high rotation measures of distant quasars with absorption systems at lower redshift. Bernet *et al.* [75] and Kronberg *et al.* [76] found that quasars with optical Mg II absorption systems with a redshift smaller than that of the quasar, indicative of an unrelated normal galaxy in the line of sight to the quasar, have a substantially higher spread in rotation measure than quasars without Mg II absorption. The inferred magnetic field strengths are similar to those in nearby spiral galaxies, leading Kronberg *et al.* [76] to the conclusion that magnetic field strengths similar to those in present day galaxies already existed a few Gyr after the big bang. The first predictions of the contribution of spiral galaxies to deep polarised radio source counts were made by Stil *et al*. [77].

## CURRENT AND FUTURE PROJECTS WITH SIGNIFICANT CANADIAN CONTENT

Canada has gained a reputation for excellence in sensitive wide field polarisation imaging. Continuing work on the DRAO deep fields [78] provides the most sensitive polarisation image of the sky to date. Canada also participates in a number of international projects that are bound to revolutionize our understanding of the origin and evolution of cosmic magnetic fields. For much of the future work on Galactic magnetism, the significant quantity is not only polarisation angle, but specifically rotation measure. All projects listed below will utilize multiple channels allowing for studies in *rotation measure synthesis*, where images may be formed at individual values of RM [79].

The Global Magneto-Ionic Medium Survey (GMIMS; principal investigator M. Wolleben, NRC) is an international project utilizing several facilities from around the world to map polarised emission across the *entire* sky from 16 cm to 1 m (300 MHz to 1.8 GHz). Specialized receivers designed and built at DRAO are being used by the participating telescopes. With observations commencing in April of 2008, the project will survey the diffuse polarised emission from the Galactic disk to the halo, at a resolution of 0.5 degrees.

Complementary to GMIMS is the Galactic Arecibo L-band Feed Array Continuum Transit Survey (GALFACTS; principal investigator A. R. Taylor, Calgary). GALFACTS is a polarisation survey with the Arecibo radio telescope that will have a sensitivity of μJy and hundreds of spectral channels. A new multi-beam cleaning technique was developed in Calgary to make high-fidelity images of compact polarised sources and diffuse emission with the seven-beam Arecibo L-Band Feed Array (ALFA). GALFACTS will extend to 32% of the sky the kind of analysis that has so far been restricted to small deep fields. The many frequency channels and large bandwidth will open the possibility to study the wavelength-dependent polarisation in much more detail than any previous survey.

The Very large Array (VLA) in New Mexico is currently being upgraded to become more than an order of magnitude more sensitive than before. Key to the upgrade of the VLA is the new central correlator that has been designed and built at DRAO. The Expanded Very Large Array will be a much more versatile instrument with a larger instantaneous bandwidth, suitable to make deep polarisation images of the sky.

Canada is a partner in the Australian Square Kilometre Array Pathfinder (ASKAP [80]). This technology demonstrator for the much larger Square Kilometre Array, scheduled to be completed by 2020, will explore wide-field imaging (30 square degree instantaneous field of view) at bandwidth of 300 MHz divided into 16000 frequency channels. Apart from the engineering challenges for this new-generation radio telescope, SKA pathfinders such as ASKAP provide new challenges in terms of image calibration and processing. Canada is expected to play a leadership role in developing techniques for wide-field polarisation imaging and calibration for both SKA pathfinders (including ASKAP) and the SKA itself. Finally, one of the five science drivers for the SKA is the origin and evolution of cosmic magnetism. Canada has demonstrated expertise on both the technical and scientific fronts defining the SKA, and we will undoubtedly continue to do so.

## ACKNOWLEDGEMENTS

The authors acknowledge support from the Natural Sciences and Engineering Research Council through the Discovery Grants program. The Canadian Galactic Plane Survey is a Canadian project with international partners and is also supported by the Natural Sciences and Engineering Research Council. The Dominion Radio Astrophysical Observatory is operated as a national facility by the National Research Council.